\documentclass[sigconf]{acmart}

\usepackage{multirow}
\usepackage{longtable}
\usepackage{algorithm}
\usepackage[noend]{algpseudocode}
\usepackage{array}
\usepackage{mdwmath}
\usepackage{mdwtab}
\usepackage{eqparbox}
\usepackage{url}
\usepackage{amsthm}
\usepackage{bbm}
\usepackage{mathtools}
\newtheorem{definition}{Definition}
\newtheorem{theorem}{Theorem}

\AtBeginDocument{%
  \providecommand\BibTeX{{%
    \normalfont B\kern-0.5em{\scshape i\kern-0.25em b}\kern-0.8em\TeX}}}

\acmConference[San Diego '20]{San Diego '20: 
The Fifth International Workshop on Fashion and KDD}{23 August 2020}{San Diego, California - USA}

\begin{document}

\title[Application of Newsboy Problem]{\LARGE An Application of Newsboy Problem in Supply Chain Optimisation of Online Fashion E-Commerce}

\author{K Chandramouli}
\authornote{Authors contributed equally to this work.}
\email{chandramouli.k@myntra.com}
\affiliation{%
  \institution{Data Science Division \\ Myntra Designs}
  \city{Bangalore}
  \state{Karnataka}
  \country{India}
}

\author{A Gopinath}
\authornotemark[1]
\email{gopinath.a@myntra.com}
\affiliation{%
  \institution{Supply Chain Management \\ Myntra Designs}
  \city{Bangalore}
  \state{Karnataka}
  \country{India}}

\author{Nachiappan Sundaram}
\authornotemark[1]
\email{nachiappan.sundaram@myntra.com}
\affiliation{%
  \institution{Data Science Division \\ Myntra Designs}
  \city{Bangalore}
  \state{Karnataka}
  \country{India}
}

\author{T Ravindra Babu} 
\email{ravindra.babu@myntra.com}
\affiliation{%
  \institution{Data Science Division \\ Myntra Designs}
  \city{Bangalore}
  \state{Karnataka}
  \country{India}
  }

\author{Chaithanya Bandi}
\email{chaithanya.bandi@myntra.com}
\affiliation{%
  \institution{Kellogg School of Management}
  \state{Illinois}
  \country{USA}
  }

\renewcommand{\shortauthors}{Chandramouli et al.}



\begin{abstract}
We describe a supply chain optimization model deployed in an online fashion e-commerce company in India called Myntra. Our model is simple, elegant and easy to put into service. The model utilizes historic data and predicts the quantity of Stock Keeping Units (SKUs) to hold so that the metrics ``Fulfilment Index" and ``Utilization Index" are optimized. We present the mathematics central to our model as well as compare the performance of our model with baseline regression based solutions.
\end{abstract}






\maketitle
\section{Introduction}
The single-item single-period newsboy problem has been extensively studied in the literature. It is still an active research problem with several extensions to real-world scenarios. In its basic formulation the problem aims to obtain a replenishment strategy for a perishable item with a stochastic demand such that the expected profit is optimized. Many extensions and modifications that introduce additional complexity to this basic formulation are proposed in the literature.

Some of the research effort on extensions of newsboy problem are focused on businesses that wish to maximize expected profit with diverse range of assumptions on demand distribution and inventory models. For example \cite{petruzzi1999pricing} assumes additive as well as multiplicative price-demand relationship in the newsboy problem to maximize the expected profit. \cite{federgruen1999combined} presents a multi-period model for inventory control with price-dependent stochastic demand. In \cite{scarf1957min}, a variation of newsboy problem is considered where the objective is to maximize the minimum profit given only the mean and variance of the demand distribution. Moreover it has been shown that the distribution that achieves this objective is close to a Poisson distribution.

Other modifications to classical newsboy problem include multi-item considerations \cite{nahmias1984efficient,lau1996newsstand} with capacity constraints for the items, multi-location models \cite{chen1990example,chang1991effect} with dependence on season. Recently deep-learning based methods have also been employed \cite{oroojlooyjadid2020applying,zhang2017assessing} to solve the newsboy problem by mitigating the dependency on demand distribution.

In the area of e-commerce \cite{ma2017news,govindarajan2017inventory} invoke newsboy problem and describe models similar to our model. However our primary focus is on fashion e-commerce. Here, as a consequence of dynamic nature of fashion and limited total quantity of a typical style/item, we face challenges like limited history of sales, frequent change of products on the platform etc. We approach these challenges with a simple choice for demand distribution based on heuristics.

In our work in fashion e-commerce we model the problem of systematic placement of inventory as an instance of newsboy problem and obtain an efficient algorithm that optimizes our business metrics ``Fulfilment Index" and ``Utilization Index". Our key contributions are summarised as follows.

\subsection*{Our contributions:}
\begin{itemize}
    \item We define and mathematically formulate our business objectives.
    \item We model the problem of optimization of our objectives as an instance of newsboy problem and solve it analytically to obtain globally optimal solution.
    \item Based on heuristic analysis of this optimal solution, we develop a SKU allocation algorithm.
    \item We establish better performance of our algorithm compared to regression based models to solve the SKU allocation problem on real world data.
\end{itemize}
\section{Problem Statement}
\label{problem_statement}
Most business organizations admit that the supply chain is the backbone of their day to day operations. So it is critical for the supply chain to be efficient and optimized.

At Myntra our supply chain outbound works as follows.
The country is divided into clusters based on the zipcode. In each cluster a local warehouse/ Forward Deployment Center (FDC) hosts a select group of products. 
An order of a product from the customer of a cluster first reaches the FDC of the cluster and gets delivered if it is available at the FDC. If the product is unavailable at the FDC the order is forwarded to the central warehouse/Big Box (BB). 

For better customer satisfaction through faster fulfillment of orders and efficient delivery it is essential that an ordered product be available at the FDC of the cluster most of the time. One way to accomplish this task is to hold a diverse range and quantity of products at the FDC and replenish them every week. But due to space and financial constraints it is infeasible to host a large number of products at each FDC of the country. Clearly there is a trade-off between these two requirements. The following definitions quantify these requirements. 

\begin{definition}[]
Fulfilment Index (FI): \\
It is defined as the ratio between number of SKUs delivered from the FDC of the cluster and number of SKUs ordered from the cluster in a week.
$$FI(cluster)= \frac{\# \text{Delivered SKUs in the week}}{\# \text{Ordered SKUs in the week}}$$
\end{definition}
\begin{definition}
Utilization Index (UI):\\
It is defined as the ratio between number of SKUs predicted for sale by the model for a given FDC in a week and number of SKUs delivered from the given FDC in the previous week.
$$UI(cluster)= \frac{\#\text{predicted SKUs for the week}}{\#\text{SKUs sold in the previous week}} $$
\end{definition}

Now from the definition of FI it is clear that the 
metric measures the efficiency of our storage. Higher values of FI signify efficient storage of high demand products and better customer service for the customers of the FDC cluster. On the other hand UI measures the space efficiency or congestion at the FDC. Moreover the ability to control this metric, accomplished by incorporating a tunable parameter $r$ in the solution, accounts for fluctuations in manpower allotment for the SKU movement. High UI for large number of weeks signify that the FDC is getting congested. In summary a FI of $100\%$ at $1$ UI is an ideal situation where every week all the items moved to FDC are sold completely.

Also observe that FI and UI are dependent metrics. For example higher values of UI account for large and diverse range of products at FDC thereby ensuring higher values of FI but with significant storage cost as well as eviction cost of unsold items.
Our objective is to come up with a solution that maximizes FI  and minimizes UI simultaneously. We achieve this objective by modeling our problem as an instance of newsboy problem.

\section{Solution}
\label{solution}
First we define approximations for the FI of a cluster and UI of a cluster at the level of SKU and 
model the reduced problem as an instance of newsboy problem. 
To begin we setup some notation. Given a SKU, denote by
\begin{align*}
     D &: \text{ Demand random variable of the given SKU }  \\
     f &: \text{ Probability density function of } D \text{ with support in } (0, \infty) \\
     F &: \text{ Cumulative distribution function of } D \\
     q &: \text{ Quantity of a given SKU to be transported to the FDC} \\
     s &: \text{ Last week sales of SKU } \\
     r &: \text{ Relative importance between FI and UI} \\
     \mathbbm{E} &: \text{ Expectation with respect to } F
\end{align*}
Similar to FI/UI of a cluster we look into FI/UI of the given SKU.
\begin{align*}
    FI(SKU)  \coloneqq &\frac{\text{\#Deliveries of the SKU in the week}}{\text{\# Orders of the SKU in the week}} \\
    = & \frac{\min\{q,D\}}{D} \\
    UI(SKU)  \coloneqq & \frac{\#\text{Predicted quantity of the SKU for the week}}{\#\text{Quantity of the SKU sold in the previous week}} \\
    = & \frac{q}{s}
\end{align*}
To achieve the objective consider the following allocation function of $q$, $a(q)$.
\begin{align*}
a(q) & \coloneqq \mathbbm{E}\left[FI(SKU)-r ~ UI(SKU)\right] \\
     & = \mathbbm{E}\left[\frac{\min\{q,D\}}{D}- r\frac{q}{s}\right]\\
     & =\mathbbm{E}\left[\frac{\min\{q,D\}}{D} \mathbbm{1}_{\{D\leq q\}}+\frac{\min\{q,D\}}{D} \mathbbm{1}_{\{D>q\}}- r\frac{q}{s}\right]\\
     & =F(q)+q\mathbbm{E}\left[\frac{1}{D} \mathbbm{1}_{\{D>q\}}\right]- r\frac{q}{s}\\
     & =F(q)+q\int^{\infty}_{q}\frac{1}{x}f(x)dx - r\frac{q}{s}\\
\end{align*}
\begin{theorem}
$q^*$ given by $\displaystyle\int^{\infty}_{q^*}\frac{1}{x}f(x)dx=\frac{r}{s}$ is the global maximizer of allocation function $a$.
\end{theorem}
\begin{proof}
Differentiating $a(q)$ we get
\begin{align*}
a'(q)& =f(q)+\int^{\infty}_{q}\frac{1}{x}f(x)dx-q\frac{f(q)}{q}- \frac{r}{s}\\
& =\int^{\infty}_{q}\frac{1}{x}f(x)dx- \frac{r}{s}\\
\end{align*}
Moreover on twice differentiating $a(q)$ we get
\begin{align*}
    a''(q)=-\frac{f(q)}{q}\leq 0.
\end{align*}
So $a'(q^*)=0$ and $a''(q*)\leq 0.$
Hence $a(q)$ is convex and $q^*$ is a global maximizer of allocation function $a.$
Now observe that for $q_1 < q_2$ 
\begin{align*}
    & \int^{\infty}_{q_1}\frac{1}{x}f(x)dx>\int^{\infty}_{q_2}\frac{1}{x}f(x)dx \\
    \iff & \int^{q_2}_{q_1}\frac{1}{x}f(x)dx>0.
\end{align*}
Therefore $a'(q)$ is strictly decreasing function of $q$. Hence $a'(q)=0$ has a unique solution. So $q^*$ is the unique global maximizer of the allocation function. 
\end{proof}

Motivated by  \cite{scarf1957min}, we employ the following heuristics as the demand of a SKU in the case of fashion e-commerce is a discrete distribution. We assume that each SKU follows a Poisson distribution. Therefore
\begin{align*}
    \int^{\infty}_{q}\frac{1}{x}f(x)dx &\approx \sum_{k \geq q}\frac{1}{k}e^{-\lambda}\frac{\lambda^k}{k!}\\
    &\approx \frac{1}{\lambda}\sum_{k \geq q}e^{-\lambda}\frac{\lambda^{k+1}}{(k+1)!}\\
    &= \frac{1-F_{\text{Poisson}}(q)}{\lambda}
\end{align*}
where $\lambda$ is the parameter of the Poisson distribution. With this heuristics we obtain the solution
\begin{align*}
    q^*=F^{-1}_{\text{Poisson}}(1-\frac{\lambda r}{s})
\end{align*}

\section{Algorithm}\label{algorithm}
\begin{algorithm}[H]
\caption{SKU allocation algorithm}
\begin{flushleft}
\textbf{Input:}\\ 
\text{Historic weekly sales of each SKU} \\    
\textbf{Output:} \\ 
\text{Quantity to allocate for each SKU}
\end{flushleft}
\begin{algorithmic}[1]
\Procedure{ALLOCATE:}{}
\For{each SKU}
\State Estimate the parameter of $F_\text{Poisson}$ for the SKU utilizing historic weekly sales of the SKU.
\State Evaluate $q_{SKU}^*=F^{-1}_{\text{Poisson}}(1-\frac{r\widehat{\lambda}_{\text{SKU}}}{s_{\text{SKU}}})$ for the SKU.
\EndFor
\State \textbf{return} $q^*_{\text{SKU}}$ for each SKU 
\EndProcedure
\end{algorithmic}
\label{alg:alloc}
\end{algorithm}
Our algorithm obtains a cluster level demand distribution $F_{\text{Poisson}}$ for each SKU by Maximum Likelihood Estimation (MLE) of the parameter of $F_{\text{Poisson}}$ for the SKU. It is easy to see that the estimator $\widehat{\lambda}_{\text{SKU}}$ given by MLE is the mean of the historical weekly sale samples of the SKU at the cluster. Observe that 
$F^{-1}_{\text{Poisson}}\left(1-\frac{r\widehat{\lambda}_{\text{SKU}}}{s_{\text{SKU}}}\right)$ is well defined if and only if $0<\left(1-\frac{r\widehat{\lambda}_{\text{SKU}}}{s_{\text{SKU}}}\right) \iff r\widehat{\lambda}_{\text{SKU}}<s_{\text{SKU}}$. Hence there is a recommendation $q^*_{\text{SKU}}$ for the SKU only under the condition that the last week sales, $s_{\text{SKU}}$, is larger than the relative importance, $r$, times the mean of sales $\widehat{\lambda}_{\text{SKU}}$ of the SKU, a intuitive thing to do, and the quantity to recommend is obtained as a quantile of the demand distribution.

\section{Experiments}
\begin{figure}[ht]
\centering
\caption{Typical SKUs in our dataset}
\includegraphics[height=1.3cm,width=1.7cm]{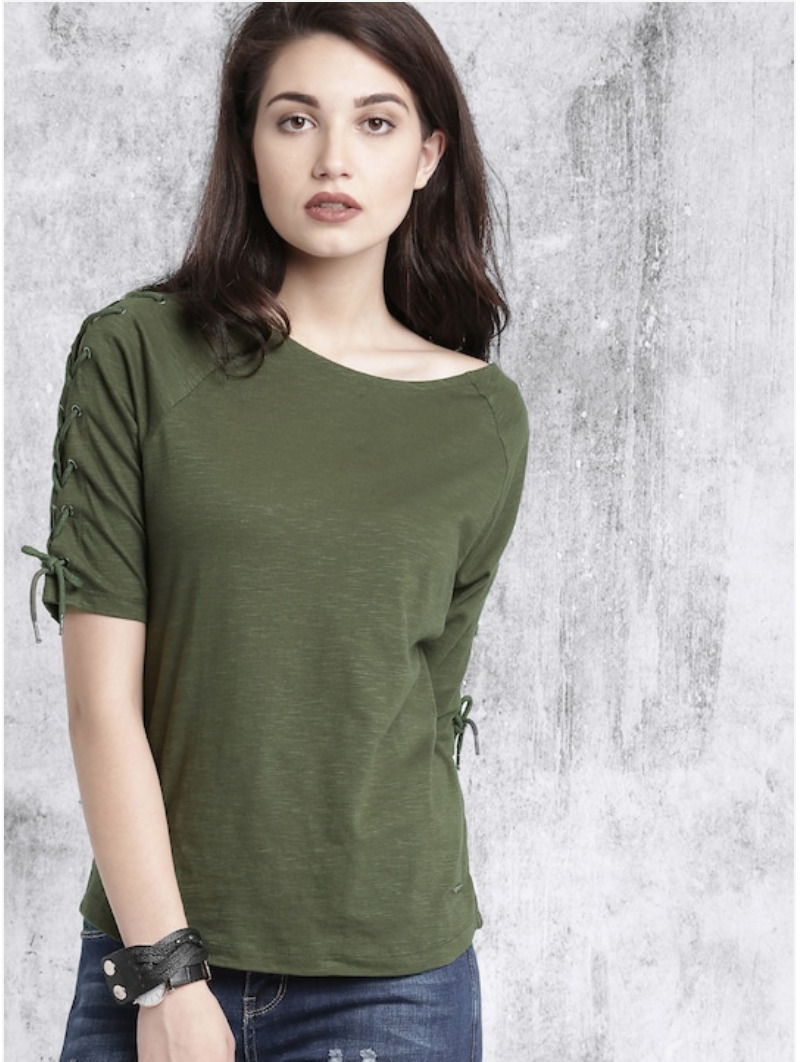}
\includegraphics[height=1.3cm,width=1.7cm]{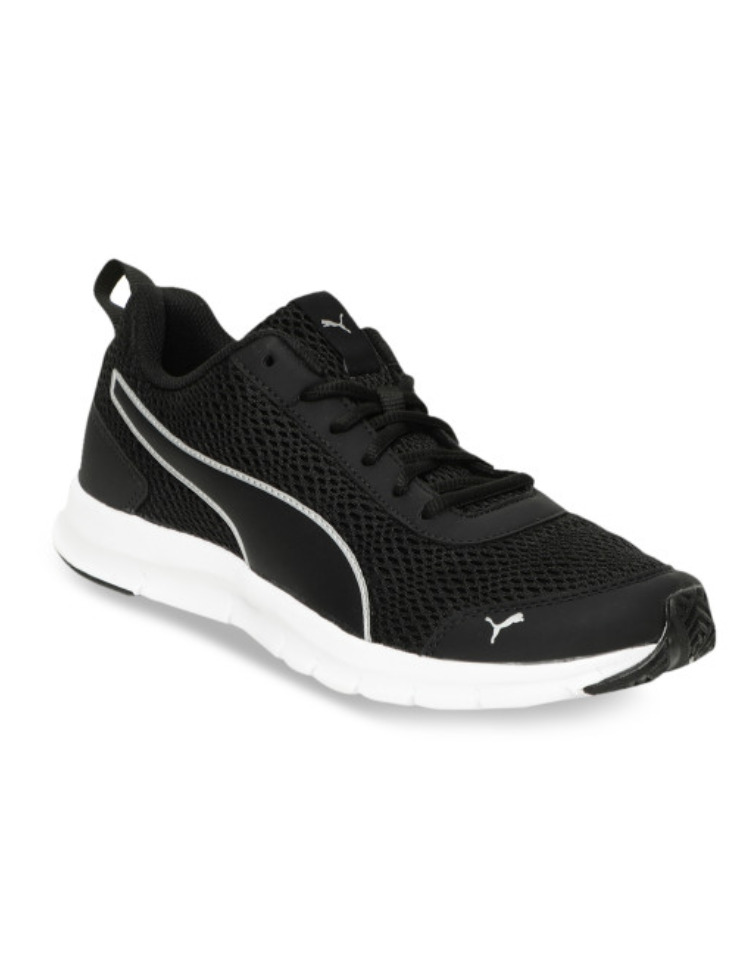}
\label{typical_SKU}
\end{figure}

To evaluate and compare our algorithm we have extensively utilized the real world data available at Myntra.com. 
First we have chosen a target week for evaluation. Then
for our algorithm \ref{alg:alloc} based on newsboy problem we prepared the dataset as described below.
Given a region/FDC/cluster (see Table \ref{newsboy_results})  and a SKU (see Figure \ref{typical_SKU}) on our platform we obtained the weekly sales of the SKU in the region for the $9$ weeks that precede the target week.
As discussed earlier in Section \ref{algorithm} we obtain the parameter of the Poisson demand distribution of the SKU in the cluster by maximum likelihood estimation. In our case the parameter estimate is the sample mean of the $9$ weekly demand samples of the SKU.
We then utilize the sample mean and weekly demand sample for the week that immediately precedes the target week,
denoted $\widehat{\lambda}_{SKU}$ and $s_{SKU}$ respectively, and evaluate step 4 of the algorithm with the choice $r=0.1$ to obtain the quantity that maximizes our allocation objective funtion. We note that the parameter $r$ allows us to trade FI for UI and vice-versa. It enables us to handle fluctuations in the implementation of the solution. The dependency of FI and UI on $r$ is summarised in Table \ref{newsboy_results} for a set of $12$ FDCs.

We compared our algorithm with a regression based model that is an earlier production model and predicts the demand of a SKU at a cluster/FDC given features of the SKU. 
The regression model is an ARMAX time series model that utilizes cluster level as well as platform level features constructed from $14$ week history of the SKU that precedes the target week 
and predicts the weekly demand of the target week.
Table \ref{features_regression_model} describes dominant features of the regression model.

We summarise our observations from the comparison of the methods here. As shown in Table \ref{comparision_results}, in almost all clusters chosen we see that both FI and UI see significant improvement over baseline method. We observe that the weekly sales data the precede the target week is also utilized (see Table \ref{features_regression_model}) in the benchmark algorithm. However the boost in performance, unlike in benchmark where the objective is mean square error, is a result of direct optimization of an objective that depends on FI and UI.

\begin{table*}[ht]
\centering
\caption{Cluster wise FI and UI metrics with different $r$-values}
\begin{tabular}{|l|l|r|l|r|l|r|l|r|l|r|}
\hline
\multicolumn{1}{|c|}{}                         & \multicolumn{2}{l|}{$r=0.025$}                                                                              & \multicolumn{2}{l|}{$r=0.05$}                                                                               & \multicolumn{2}{l|}{$r=0.1$}                                                                                & \multicolumn{2}{l|}{$r=0.2$}                                                                                & \multicolumn{2}{l|}{$r=0.4$}                                                                                \\ \cline{2-11} 
\multicolumn{1}{|c|}{\multirow{-2}{*}{\textbf{Region}}} & \textbf{FI}                                                 & \multicolumn{1}{l|}{\textbf{UI}}                         & \textbf{FI}                                                 & \multicolumn{1}{l|}{\textbf{UI}}                         & \textbf{FI}                                                 & \multicolumn{1}{l|}{\textbf{UI}}                         & \textbf{FI}                                                 & \multicolumn{1}{l|}{\textbf{UI}}                         & \textbf{FI}                                                 & \multicolumn{1}{l|}{\textbf{UI}}                         \\ \hline
FDC\_1                                      &  {  44\%} &  {  3.07} &  {  44\%} &  {  2.92} &  {  36\%} &  {  0.85} &  {  35\%} &  {  0.67} &  {  32\%} &  {  0.57} \\ \hline
FDC\_2                                      &  {  70\%} &  {  2.86} &  {  69\%} &  {  2.64} &  {  67\%} &  {  0.97} &  {  63\%} &  {  0.80} &  {  58\%} &  {  0.64} \\ \hline
FDC\_3                                        &  {  38\%} &  {  3.33} &  {  38\%} &  {  3.22} &  {  31\%} &  {  0.83} &  {  29\%} &  {  0.67} &  {  28\%} &  {  0.58} \\ \hline
FDC\_4                                          &  {  68\%} &  {  2.53} &  {  67\%} &  {  2.30} &  {  62\%} &  {  0.98} &  {  58\%} &  {  0.82} &  {  53\%} &  {  0.64} \\ \hline
FDC\_5                                      &  {  44\%} &  {  3.37} &  {  44\%} &  {  3.18} &  {  36\%} &  {  0.88} &  {  34\%} &  {  0.70} &  {  31\%} &  {  0.58} \\ \hline
FDC\_6                                      &  {  61\%} &  {  3.28} &  {  60\%} &  {  3.04} &  {  52\%} &  {  0.98} &  {  50\%} &  {  0.82} &  {  46\%} &  {  0.66} \\ \hline
FDC\_7                                        &  {  42\%} &  {  3.11} &  {  42\%} &  {  2.97} &  {  35\%} &  {  0.83} &  {  33\%} &  {  0.67} &  {  31\%} &  {  0.57} \\ \hline
FDC\_8                                        &  {  40\%} &  {  3.10} &  {  40\%} &  {  2.97} &  {  34\%} &  {  0.83} &  {  32\%} &  {  0.66} &  {  31\%} &  {  0.57} \\ \hline
FDC\_9                                       &  {  54\%} &  {  2.59} &  {  53\%} &  {  2.38} &  {  45\%} &  {  0.90} &  {  42\%} &  {  0.71} &  {  38\%} &  {  0.58} \\ \hline
FDC\_10                                      &  {  34\%} &  {  3.14} &  {  35\%} &  {  3.03} &  {  28\%} &  {  0.75} &  {  27\%} &  {  0.60} &  {  25\%} &  {  0.52} \\ \hline
FDC\_11                                         &  {  71\%} &  {  2.86} &  {  70\%} &  {  2.63} &  {  69\%} &  {  0.98} &  {  65\%} &  {  0.81} &  {  61\%} &  {  0.65} \\ \hline
FDC\_12                                         &  {  46\%} &  {  2.84} &  {  46\%} &  {  2.67} &  {  37\%} &  {  0.86} &  {  35\%} &  {  0.69} &  {  32\%} &  {  0.58} \\ \hline
\end{tabular}
\label{newsboy_results}
\end{table*}



\begin{table*}[ht]
\centering
\caption{Comparison of cluster wise FI and UI metrics for both models}
\begin{tabular}{|l|l|l|r|r|}
\hline
\multirow{2}{*}{\textbf{Region}} & \multicolumn{2}{l|}{\textbf{Regression Model}} & \multicolumn{2}{l|}{\textbf{Our Algorithm}}                              \\ \cline{2-5} 
                                 & \textbf{FI}         & \textbf{UI}         & \multicolumn{1}{l|}{\textbf{FI}} & \multicolumn{1}{l|}{\textbf{UI}} \\ \hline
FDC\_1                        & 25\%                 & 1.24                    & \textbf{36\%}                              & \textbf{0.85}                                 \\ \hline
FDC\_2                        & 47\%                 & 1.98                    & \textbf{67\%}                              & \textbf{0.97}                                 \\ \hline
FDC\_3                          & 18\%                 & 0.81                    & \textbf{31\%}                              & \textbf{0.83}                                 \\ \hline
FDC\_4                           & 54\%                 & 2.06                    &\textbf{62\%}                              & \textbf{0.98}                                 \\ \hline
FDC\_5                         & 21\%                 & 1.08                    & \textbf{36\%}                              & \textbf{0.88}                                 \\ \hline
FDC\_6                        & 41\%                 & 2.10                    & \textbf{52\%}                              & \textbf{0.98}                                 \\ \hline
FDC\_7                           & 21\%                 & 0.70                    & \textbf{35\%}                              & \textbf{0.83}                                 \\ \hline
FDC\_8                         & 21\%                 & 0.84                    & \textbf{34\%}                              & 0.83                                 \\ \hline
FDC\_9                         & 32\%                 & 0.83                    & \textbf{45\%}                              & \textbf{0.90}                                 \\ \hline
FDC\_10                        & 15\%                 & 0.75                    & \textbf{28\%}                              & \textbf{0.75}                                 \\ \hline
FDC\_11                           & 48\%                 & 1.99                    &\textbf{69\%}                              & \textbf{0.98}                                 \\ \hline
FDC\_12                           & 33\%                 & 1.81                    & \textbf{37\%}                              & \textbf{0.86}                                 \\ \hline
\end{tabular}
\label{comparision_results}
\end{table*}

\begin{table*}[ht]
\caption{Dominant features of regression based model}
\begin{tabular}{|l|l|}
\hline
\textbf{Features}                & \textbf{Description}                                                                       \\ \hline
atc\_count                       & Add to cart count of the SKU                                                                         \\ \hline
mrp\_range                       & Article retail price classification                                                                 \\ \hline
inventory\_refill\_days          & Average days to refill the inventory from stock out                                        \\ \hline

t\_1:12\_quantity                   & Weekly sales in the 12 weeks that precede the target week                                                           \\ \hline

three\_w\_max\_units             & Last three weeks maximum inventory in units                                                \\ \hline
three\_w\_min\_units             & Last three weeks minimum inventory in units                                                \\ \hline
sku\_selling\_rt                     & Ratio between cluster SKU sales and cluster style sales                                    \\ \hline
sku\_style\_rt                   & Ratio between overall SKU sales and overall style sales                                    \\ \hline
catalog\_live\_week\_no\_diff    &  Age of the product                                                                                  \\ \hline
wl\_count                        & Wish list count                                                                            \\ \hline
\end{tabular}
\label{features_regression_model}
\end{table*}

\section{Conclusion}
We have modeled our business problem as an instance of newsboy problem and solved it. Based on heuristic analysis of this solution we proposed a SKU allocation algorithm. We have compared and shown that our algorithm has superior business metrics compared to regression based benchmark.
\bibliographystyle{ACM-Reference-Format}
\bibliography{sample-base}


\end{document}